\documentclass{elsart}
\usepackage{epsfig}
\begin{document}
\runauthor{Cicero, Caesar and Vergil}
\begin{frontmatter}
\title{Nuclear equation of state at high baryonic density and compact star constraints}
\vskip -0.7cm

\author[VECC]{D.N. Basu\thanksref{Z}},
\vskip -0.25cm

\address[VECC]{Variable  Energy  Cyclotron  Centre, 1/AF Bidhan Nagar, Kolkata 700 064, India }
\thanks[Z]{E-mail:dnb@veccal.ernet.in}

and
\author[SINP]{P. Roy Chowdhury\thanksref{X}} and
\author[SINP,VCU]{C. Samanta\thanksref{Y}}

\vskip -0.25cm
\address[SINP]{Saha Institute of Nuclear Physics, 1/AF Bidhan Nagar, Kolkata 700 064, India }

\vskip -0.25cm
\address[VCU]{Physics Department, Virginia Commonwealth University, Richmond, VA 23284-2000, U.S.A.}

\thanks[X]{E-mail:partha.roychowdhury@saha.ac.in}
\thanks[Y]{E-mail:chhanda.samanta@saha.ac.in}

\vskip -0.4cm
\begin{abstract}
\vskip -0.4cm

      A mean field calculation is carried out to obtain the equation of state (EoS) of nuclear matter from a density dependent M3Y interaction (DDM3Y). The energy per nucleon is minimized to obtain ground state of the symmetric nuclear matter (SNM). The constants of density dependence of the effective interaction are obtained by reproducing the saturation energy per nucleon and the saturation density of SNM. The energy variation of the exchange potential is treated properly in the negative energy domain of nuclear matter. The EoS of SNM, thus obtained, is not only free from the superluminosity problem but also provides excellent estimate of nuclear incompressibility. The EoS of asymmetric nuclear matter is calculated by adding to the isoscalar part, the isovector component of M3Y interaction. The SNM and pure neutron matter EoS are used to calculate the nuclear symmetry energy which is found to be consistent with that extracted from the isospin diffusion in heavy-ion collisions at intermediate energies. The $\beta$ equilibrium proton fraction calculated from the symmetry energy and related theoretical findings are consistent with the constraints derived from the observations on compact stars. 

\end{abstract}
\vskip -0.2cm
\noindent
\vskip -0.7cm
\end{frontmatter}

\section{Introduction}

      The investigation of constraints for the high baryonic density behaviour of nuclear matter has recently received new impetus with the plans to construct a  new accelerator facility (FAIR) at GSI Dramstadt. The stiffness of a nuclear equation of state (EoS) is characterised by nuclear incompressibility \cite{Ba80} which can be extracted experimentally. Nuclear incompressibility \cite{Sa89,Sa90} also determines the velocity of sound in nuclear medium for predictions of shock wave generation and propagation. The EoS is of fundamental importance in the theories of nucleus-nucleus collisions at energies where the nuclear incompressibility $K_0$ comes into play as well as in the theories of supernova explosions \cite{Be88}. A widely used experimental method is the determination of the nuclear incompressibility from the observed giant monopole resonances (GMR) \cite{Sh88,Yo05}. Other recent experimental determinations are based upon the production of hard photons in heavy ion collisions \cite{Sc96} and from isoscalar giant dipole resonances (ISGDR) \cite{Lu04,Yo04,Ga04}. From the experimental data of isoscalar giant monopole resonance (ISGMR) conclusion can be drawn that $K_0\approx$ 240 $\pm$ 20 MeV \cite{Sh06}. The general theoretical observation is that the non-relativistic \cite{Co04} and the relativistic \cite{Br90} mean field models \cite{CG04} predict for the bulk incompressibility for the SNM, $K_0$, values which are significantly different from one another, {\it viz.} $\approx$ 220-235 MeV and $\approx$ 250-270 MeV respectively. Theoretical EoS for the SNM that predict higher $K_0$ values $\approx$ 300 MeV are often called `stiff' EoS whereas those EoS which predict smaller $K_0$ values $\approx$ 200 MeV are termed as `soft' EoS.   

      The nuclear symmetry energy (NSE) is an important quantity in the equation of state of isospin asymmetric nuclear matter. This currently unknown quantity plays a key role to the understanding of the structure of systems as diverse as the neutron rich nuclei and neutron stars and it enters as an input to the heavy ion reactions \cite{Li98,Li02}. In general, they can be broadly classified into two different forms. One, where the NSE increases monotonically with increasing density (`stiff' dependence) \cite{Ch05} and the other, where the NSE increases initially up to normal nuclear density or beyond and then decreases at higher densities (`soft dependence') \cite{Ba02}. Determination of the exact form of the density dependence of the NSE is rather important for studying the structure of neutron rich nuclei, and studies relevant to astrophysical problems, such as the structure of neutron stars and the dynamics of supernova collapse \cite{Ba01}. `Stiff' density dependence of the NSE is predicted to lead a large neutron skin thickness compared to a `soft' dependence and can result in rapid cooling of a neutron star, and a larger neutron star radius, compared to a soft density dependence. A somewhat `stiff' EoS of nuclear matter need not lead to a `stiff' density dependence of the NSE. Modern constraints from mass and mass-radius-relation measurements require a stiff EoS at high densities whereas flow data from heavy-ion collisions seem to disfavour too stiff behavior of the EoS. As a compromise hybrid EoS \cite{gr06,ba07,bu08} with a smooth transition at high density to quark matter are being proposed.  

      In view of rather large differences between the various calculations of the NSE present even at the subsaturation densities, the question arises whether one can obtain empirical constraints from finite nuclei. Since the degree of isospin diffusion in heavy-ion collisions at intermediate energies is affected by the stiffness of the NSE, these reactions, therefore, can also provide constraints on the low energy behaviour of the NSE \cite{Ch05}. However, the high density behaviour remains largely undetermined since hardly any data on the simultaneous measurements of masses and corresponding radii of neutron stars exist whereas they can be obtained theoretically by solving Tolman-Oppenheimer-Volkov equation. However there exist indirect indications such as the neutron star cooling process. Recently search for the experimental signatures of the moderately high density behaviour of the NSE has been proposed \cite{Ba02} theoretically using several sensitive probes suh as the $\pi^-$ to $\pi^+$ ratio, tranverse collective flow and its excitation function as well as the neutron-proton differential flow. 

      In the present work, we apply recently discovered astrophysical bounds of high density behaviour of nuclear matter from compact star cooling phenomenology and also show that the theoretical description of nuclear matter using the density dependent M3Y-Reid-Elliott effective interaction \cite{Be77,Sa79} gives a value of nuclear incompressibility which is in excellent agreement with values extracted from experiments. The velocity of sound does not become superluminous since the energy dependence is treated properly for the negative energy domain of nuclear matter. It also provides a symmetry energy that is consistent with the empirical value extracted from the measured masses of nuclei. The microscopic proton-nucleus interaction potential is obtained by folding the density of the nucleus with DDM3Y effective interaction whose density dependence is determined completely from the nuclear matter calculations. The quantum mechanical tunneling probability is calculated within the WKB framework using these nuclear potentials. These calculations provide reasonable estimates for the observed proton radioactivity lifetimes. Along with earlier works using the same formalism, present work provides a unified description of radioactivity, scattering, nuclear EoS and NSE. 

\section{The nuclear equation of state for symmetric nuclear matter}

      In the present work, density dependence of the effective interaction, DDM3Y, is completely determined from nuclear matter calculations. The equilibrium density of the nuclear matter is determined by minimizing the energy per nucleon. In contrast to our earlier calculations for the nuclear EoS where the energy dependence of the zero range potential was treated as fixed at a value corresponding to the equilibrium energy per nucleon $\epsilon_0$ \cite{BCS06} and assumed to vary negligibly with $\epsilon$ inside nuclear fluid, in the present calculations the energy variation of the zero range potential is treated more accurately by allowing it to vary freely but only with the kinetic energy part $\epsilon^{kin}$ of the energy per nucleon $\epsilon$ over the entire range of $\epsilon$. This is not only more plausible, but also yields excellent result for the incompressibility $K_0$ of the SNM which no more suffers from the superluminosity problem. 

      The constants of density dependence are determined by reproducing the saturation conditions. It is worthwhile to mention here that due to attractive character of the M3Y forces the saturation condition for cold nuclear matter is not fulfilled. However, the realistic description of nuclear matter properties can be obtained with this density dependent M3Y effective interaction. Therefore, the constants of density dependence have been obtained by reproducing the saturation energy per nucleon and the saturation nucleonic density of the cold SNM. Based on the Hartree or mean field assumption and using the DDM3Y interaction, the expression for the energy per nucleon for symmetric nuclear matter $\epsilon$ is given by  

\begin{equation}
 \epsilon = [\frac{3\hbar^2k_F^2}{10m}] + [\frac{\rho J_{v00} C (1 - \beta\rho^n)}{2}]  
\label{seqn1}
\end{equation}
\noindent
where Fermi momentum $k_F = (1.5\pi^2\rho)^{\frac{1}{3}}$,  $m$ is the nucleonic mass equal to 938.91897 MeV/c$^2$ and $J_{v00}$ represents the volume integral of the isoscalar part of the M3Y interaction supplemented by the zero-range potential having the form    
 
\begin{eqnarray}
 && J_{v00} = J_{v00}(\epsilon^{kin})=\int \int \int t_{00}^{M3Y}(s, \epsilon) d^3s \nonumber \\ =&& 7999\frac{4\pi}{4^3} - 2134\frac{4\pi}{2.5^3} + J_{00} (1 - \alpha\epsilon^{kin})
 ~{\rm where}~J_{00}=-276~{\rm MeV~fm^3}. 
\label{seqn2}
\end{eqnarray}
\noindent
where $\epsilon^{kin} = \frac{3\hbar^2k_F^2}{10m}$ is the kinetic energy part of the energy per nucleon $\epsilon$ given by Eq.(1).

      The isoscalar $t_{00}^{M3Y}$ and the isovector $t_{01}^{M3Y}$ components of M3Y interaction potentials \cite{Sa79,BCS05} supplemented by zero range potentials are given by $t_{00}^{M3Y}(s, \epsilon) = 7999\frac{\exp( - 4s)}{4s} - 2134\frac{\exp( - 2.5s)}{2.5s} - 276 (1 - \alpha\epsilon)\delta(s)$ and $t_{01}^{M3Y}(s, \epsilon) = -4886\frac{\exp( - 4s)}{4s} + 1176\frac{\exp( - 2.5s)}{2.5s} + 228 (1 - \alpha\epsilon)\delta(s)$ respectively, where the energy dependence parameter $\alpha$=0.005/MeV. The DDM3Y effective NN interaction is given by $v_{0i}(s,\rho, \epsilon) = t_{0i}^{M3Y}(s, \epsilon) g(\rho)$ where the density dependence $g(\rho) = C (1 - \beta \rho^n)$ and the constants $C$ and $\beta$ of density dependence have been obtained from the saturation condition $\frac{\partial\epsilon}{\partial\rho} = 0$ at $\rho = \rho_{0}$ and $\epsilon = \epsilon_{0}$ where $\rho_{0}$ and $\epsilon_{0}$ are the saturation density and the saturation energy per nucleon respectively.
Eq.(1) can be differentiated with respect to $\rho$ to yield equation  

\begin{equation}
 \frac{\partial\epsilon}{\partial\rho} = [\frac{\hbar^2k_F^2}{5m\rho}] + \frac{J_{v00} C}{2} [1 - (n+1)\beta\rho^n] 
-\alpha J_{00} C [1 - \beta\rho^n]  [\frac{\hbar^2k_F^2}{10m}].
\label{seqn3}
\end{equation}
\noindent
The equilibrium density of the cold SNM is determined from the saturation condition. Then Eq.(1) and Eq.(3) with the saturation condition $\frac{\partial\epsilon}{\partial\rho} = 0$ can be solved simultaneously for fixed values of the saturation energy per nucleon $\epsilon_0$ and the saturation density $\rho_{0}$ of the cold SNM to obtain the values of $\beta$ and $C$. The constants of density dependence $\beta$ and $C$, thus obtained, are given by 

\begin{equation}
 \beta = \frac{[(1-p)+(q-\frac{3q}{p})]\rho_{0}^{-n}}{[(3n+1)-(n+1)p+(q-\frac{3q}{p})]}
\label{seqn4}
\end{equation} 
\noindent

\begin{equation}
 {\rm where}~~~~p = \frac{[10m\epsilon_0]}{[\hbar^2k_{F_0}^2]},~q=\frac{2\alpha\epsilon_0J_{00}}{J^0_{v00}}
\label{seqn5}
\end{equation} 
\noindent
where $J^0_{v00} = J_{v00}(\epsilon^{kin}_0)$ which means $J_{v00}$ at $\epsilon^{kin}=\epsilon^{kin}_0$, the kinetic energy part of the saturation energy per nucleon of SNM,  $k_{F_0} = [1.5\pi^2\rho_0]^{1/3}$ and 

\begin{equation}
 C = -\frac{[2\hbar^2k_{F_0}^2] }{ 5mJ^0_{v00} \rho_0 [1 - (n+1)\beta\rho_0^n -\frac{q\hbar^2k_{F_0}^2 (1-\beta\rho_0^n)}{10m\epsilon_0}]},
\label{seqn6}
\end{equation} 
\noindent
respectively. It is quite obvious that the constants of density dependence $C$ and $\beta$ obtained by this method depend on the saturation energy per nucleon $\epsilon_0$, the saturation density $\rho_{0}$, the index $n$ of the density dependent part and on the strengths of the M3Y interaction through the volume integral $J^0_{v00}$. 

\section{The incompressibility of symmetric nuclear matter}

      The incompressibility or the compression modulus of symmetric nuclear matter, which is a measure of the curvature of an EoS at saturation density and defined as $k_F^2\frac{\partial^2\epsilon}{\partial{k_F^2}} \mid_{k_F=k_{F_0}}$, measures the stiffness of an EoS. The $\frac{\partial^2\epsilon}{\partial\rho^2}$ is given by

\begin{eqnarray}
 \frac{\partial^2\epsilon}{\partial\rho^2} =&& [-\frac{\hbar^2k_F^2}{15m\rho^2}] - [\frac{J_{v00} C n(n+1) \beta\rho^{n-1}}{2}] \nonumber \\
&&- \alpha J_{00} C [1-(n+1)\beta\rho^n] [\frac{\hbar^2k_F^2}{5m\rho}] 
+ [\frac{\alpha J_{00} C (1-\beta\rho^n)\hbar^2k_F^2}{30m\rho}]
\label{seqn7}
\end{eqnarray}
\noindent
and therefore the incompressibility $K_0$ of the cold SNM which is defined as   
  
\begin{equation}
 K_0 = k_F^2\frac{\partial^2\epsilon}{\partial{k_F^2}} \mid_{k_F=k_{F_0}} = 9\rho^2\frac{\partial^2\epsilon}{\partial\rho^2} \mid_{\rho=\rho_0}
\label{seqn8}
\end{equation}
\noindent
can be theoretically obtained as

\begin{eqnarray}
 K_0 &&= [-\frac{3\hbar^2k_{F_0}^2}{5m}] - [\frac{9 J^0_{v00} C n(n+1) \beta\rho_0^{n+1}}{2}] \nonumber \\
&&- 9\alpha J_{00} C [1-(n+1)\beta\rho_0^n] [\frac{\rho_0\hbar^2k_{F_0}^2}{5m}] 
+ [\frac{3\rho_0\alpha J_{00} C (1-\beta\rho_0^n)\hbar^2k_{F_0}^2}{10m}].
\label{seqn9}
\end{eqnarray} 
\noindent

      The calculations are performed using the values of the saturation density $\rho_0$=0.1533 fm$^{-3}$ \cite{Sa89} and the saturation energy per nucleon $\epsilon_0$=-15.26 MeV \cite{CSB05,CB06} for the SNM obtained from the co-efficient of the volume term of Bethe-Weizs\"acker mass formula \cite{Bw35,Bw36} which is evaluated by fitting the recent experimental and estimated atomic mass excesses from Audi-Wapstra-Thibault atomic mass table \cite{Au03} by minimizing the mean square deviation incorporating correction for the electronic binding energy \cite{Lu03}. In a similar recent work, including surface symmetry energy term, Wigner term, shell correction and proton form factor correction to Coulomb energy also, $a_v$ turns out to be 15.4496 MeV \cite{Ro06} ($a_v$ =14.8497 MeV when $A^0$ and $A^{1/3}$ terms are also included). Using the usual values of $\alpha$=0.005 MeV$^{-1}$ for the parameter of energy dependence of the zero range potential and $n$=2/3, the values obtained for the constants of density dependence $C$ and $\beta$ and the SNM incompressibility $K_0$ are 2.2497, 1.5934 fm$^2$ and 274.7 MeV respectively. The saturation energy per nucleon is the volume energy coefficient and the value of -15.26$\pm$0.52 MeV covers, more or less, the entire range of values obtained for $a_v$ for which now the values of $C$=2.2497$\pm$0.0420, $\beta$=1.5934$\pm$0.0085 fm$^2$ and the SNM incompressibility $K_0$=274.7$\pm$7.4 MeV.  

      The theoretical estimate $K_0$ of the incompressibility of infinite SNM obtained from present approach using DDM3Y is about 270 MeV. The theoretical estimate of $K_0$ from the refractive $\alpha$-nucleus scattering is about 240-270 MeV \cite{Kh97,Kh07} and that by infinite nuclear matter model (INM) \cite{Sa99} claims a well defined and stable value of $K_0=288\pm20$ MeV and present theoretical estimate is in reasonably close agreement with the value obtained by INM which rules out any values lower than 200 MeV. Present estimate for the incompressibility $K_0$ of the infinite SNM is in good agreement with the experimental value of $K_0=300\pm25$ MeV obtained from the giant monopole resonance (GMR) \cite{Sh88} and with the the recent experimental determination of $K_0$ based upon the production of hard photons in heavy ion collisions which led to the experimental estimate of $K_0=290\pm50$ MeV \cite{Sc96}. However, the experimental values of $K_0$ extracted from the isoscalar giant dipole resonance (ISGDR) are claimed to be smaller \cite{Ga04}. The present non-relativistic mean field model estimate for the nuclear incompressibility $K_0$ for SNM using DDM3Y interaction is rather close to the theoretical estimates obtained using relativistic mean field models and close to the lower limit of the older experimental values \cite{Sh88} and close to the upper limit of the recent values \cite{Yo05} extracted from experiments.

      Considering the status of experimental determination of the SNM incompressibility from data on the compression modes ISGMR and ISGDR of nuclei it can be inferred \cite{Sh06} that due to violations of self consistency in HF-RPA calculations of the strength functions of giant resonances result in shifts in the calculated values of the centroid energies which may be larger in magnitude than the current experimental uncertainties. In fact, the prediction of $K_0$ lying in the range of 210-220 MeV were due to the use of a not fully self-consistent Skyrme calculations \cite{Sh06}. Correcting for this drawback, Skyrme parmetrizations of SLy4 type predict $K_0$ values in the range of 230-240 MeV \cite{Sh06}. Moreover, it is possible to build bona fide Skyrme forces so that the SNM incompressibility is close to the relativistic value, namely 250-270 MeV. Therefore, from the ISGMR experimental data the conclusion can be drawn that $K_0\approx$ 240 $\pm$ 20 MeV. The ISGDR data tend to point to lower values \cite{Lu04,Yo04,Ga04} for $K_0$. However, there is consensus that the extraction of $K_0$ is in this case more problematic for various reasons. In particular, the maximum cross-section for ISGDR decreases very strongly at high excitation energy and may drop below the current experimental sensitivity for excitation energies \cite{Sh06} above 30 and 26 MeV for $^{116}$Sn and $^{208}$Pb, respectively. The present value of 274.7$\pm$7.4 MeV for the incompressibility $K_0$ of SNM obtained using DDM3Y interaction is, therefore, an excellent theoretical result.       
 
      The constant of density dependence $\beta$=1.5934$\pm$0.0085 fm$^2$, which has the dimension of cross section for $n$=2/3, can be interpreted as the isospin averaged effective nucleon-nucleon interaction cross section in ground state symmetric nuclear medium. For a nucleon in ground state nuclear matter $k_F\approx$ 1.3 fm$^{-1}$ and $q_0 \sim \hbar k_F c \approx$ 260 MeV and the present result for the `in medium' effective cross section is reasonably close to the value obtained from a rigorous Dirac-Brueckner-Hartree-Fock \cite{Sa06} calculations corresponding to such $k_F$ and $q_0$ values which is $\approx$ 12 mb. Using the value of the constant of density dependence $\beta$=1.5934$\pm$0.0085 fm$^2$ corresponding to the standard value of the parameter $n$=2/3 along with the nucleonic density of 0.1533 fm$^{-3}$, the value obtained for the nuclear mean free path $\lambda$ is about 4 fm which is in excellent agreement \cite{Si83} with that obtained using another method. 

\section{The nuclear equation of state for asymmetric nuclear matter}

      The EoS for asymmetric nuclear matter is calculated by adding to the isoscalar part, the isovector component \cite{La62} of M3Y interaction \cite{Sa83} that do not contribute to the EoS of SNM. The EoS for SNM and pure neutron matter (PNM) are then used to calculate NSE. In this section, implications for the density dependence of this NSE in case of neutron stars are discussed, and also possible constraints on the density dependence obtained from finite nuclei are compared.

      Assuming interacting Fermi gas of neutrons and protons, with isospin asymmetry $X= \frac{\rho_n - \rho_p} {\rho_n + \rho_p},~~~~\rho = \rho_n+\rho_p,$ where $\rho_n$, $\rho_p$ and $\rho$ are the neutron, proton and nucleonic densities respectively, the energy per nucleon for isospin asymmetric nuclear matter can be derived as

\begin{eqnarray}
 &&\epsilon(\rho,X) = [\frac{3\hbar^2k_F^2}{10m}] F(X) + (\frac{\rho J_v C}{2}) (1 - \beta\rho^n) \nonumber \\
 &&= [\frac{3\hbar^2k_F^2}{10m}] F(X) - [\frac{\hbar^2k_{F_0}^2}{5m}] [\frac{\rho}{\rho_{0}}] [\frac{J_v}{J^0_{v00}}] {\big [}\frac{ (1 - \beta\rho^n)}{1 - (n+1)\beta\rho_{0}^n-\frac{q\hbar^2k_{F_0}^2 (1-\beta\rho_0^n)}{10m\epsilon_0}}{\big ]}
\label{seqn10}
\end{eqnarray}
\noindent
where $k_F= (1.5\pi^2\rho)^{\frac{1}{3}}$ which is equal to Fermi momentum in case of SNM, the kinetic energy per nucleon $\epsilon^{kin} = [\frac{3\hbar^2k_F^2}{10m}] F(X)$ with $F(X) = [\frac{(1+X)^{5/3} + (1-X)^{5/3}}{2}]$ and $J_v=J_{v00} + X^2 J_{v01}$ and $J_{v01}$ represents the volume integral of the isovector part of the M3Y interaction supplemented by the zero-range potential having the form    
 
\begin{eqnarray}
 &&J_{v01} = J_{v01}(\epsilon^{kin}) = \int \int \int t_{01}^{M3Y}(s, \epsilon) d^3s \nonumber \\
=&& -4886\frac{4\pi}{4^3} + 1176\frac{4\pi}{2.5^3} + J_{01} (1 - \alpha\epsilon^{kin}) 
~{\rm where}~J_{01}=228~{\rm MeV~fm^3},
\label{seqn11}
\end{eqnarray}
\noindent 

\begin{equation}
 \frac{\partial\epsilon}{\partial\rho} = [\frac{\hbar^2k_F^2}{5m\rho}] F(X) +\frac{J_v C}{2} [1 - (n+1)\beta\rho^n] 
 -\alpha J C [1 - \beta\rho^n]  [\frac{\hbar^2k_F^2}{10m}] F(X)
\label{seqn12}
\end{equation}
\noindent
where $J=J_{00}+X^2 J_{01}$ and

\begin{eqnarray}
 && \frac{\partial^2\epsilon}{\partial\rho^2} = [-\frac{\hbar^2k_F^2}{15m\rho^2}] F(X) - [\frac{J_v C n(n+1) \beta\rho^{n-1}}{2}] \nonumber \\
&&- \alpha J C [1-(n+1)\beta\rho^n] [\frac{\hbar^2k_F^2}{5m\rho}] F(X) + [\frac{\alpha J C (1-\beta\rho^n)\hbar^2k_F^2}{30m\rho}] F(X).
\label{seqn13}
\end{eqnarray}
\noindent

      The pressure $P$ and the energy density $\varepsilon$ of nuclear matter with isospin asymmetry $X$ are given by  

\begin{eqnarray}
 P = && \rho^2 \frac{\partial\epsilon}{\partial\rho} =  [\rho\frac{\hbar^2k_F^2}{5m}] F(X)+ \rho^2\frac{J_v C}{2} [1 - (n+1)\beta\rho^n] \nonumber \\
&&- \rho^2\alpha J C [1 - \beta\rho^n]  [\frac{\hbar^2k_F^2}{10m}] F(X) ,
\label{seqn14}
\end{eqnarray} 
\noindent
and

\begin{equation}
 \varepsilon = \rho (\epsilon + m c^2) = \rho [(\frac{3\hbar^2k_F^2}{10m}) F(X) + (\frac{\rho J_v C}{2}) (1 - \beta\rho^n) + m c^2], 
\label{seqn15}
\end{equation} 
\noindent
respectively, and thus the velocity of sound $v_s$ in nuclear matter with isospin asymmetry $X$ is given by 

\begin{equation}
 \frac{v_s}{c} = \sqrt{\frac{\partial P}{\partial\varepsilon}} =\sqrt{\frac{[2\rho\frac{\partial\epsilon}{\partial\rho} + \rho^2\frac{\partial^2\epsilon}{\partial\rho^2}]} {[\epsilon + m c^2 + \rho\frac{\partial\epsilon}{\partial\rho}]}}.
\label{seqn16}
\end{equation} 
\noindent
The incompressibilities for isospin asymmetric nuclear matter are evaluated at saturation densities $\rho_s$ with the condition $\frac{\partial\epsilon}{\partial\rho}=0$ which corresponds to vanishing pressure. The incompressibility $K_s$ for isospin asymmetric nuclear matter is therefore expressed as 

\begin{eqnarray}
 &&K_s = -\frac{3\hbar^2k_{F_s}^2}{5m} F(X) - \frac{9 J^s_v C n(n+1) \beta\rho_s^{n+1}}{2} \\
 &&- 9\alpha J C [1-(n+1)\beta\rho_s^n] [\frac{\rho_s\hbar^2k_{F_s}^2}{5m}] F(X) 
+ [\frac{3\rho_s\alpha J C (1-\beta\rho_s^n)\hbar^2k_{F_s}^2}{10m}] F(X). \nonumber
\label{seqn17}
\end{eqnarray} 
\noindent
Here $k_{F_s}$ means that the $k_F$ is evaluated at saturation density $\rho_s$. $J^s_v=J^s_{v00} + X^2 J^s_{v01}$ is the $J_v$ at $\epsilon^{kin}=\epsilon^{kin}_s$ which is the kinetic energy part of the saturation energy per nucleon $\epsilon_s$ and $J=J_{00} + X^2 J_{01}$.

      In Table-1 incompressibility of isospin asymmetric nuclear matter  $K_s$ as a function of the isospin asymmetry parameter $X$, using the usual value of $n$=2/3 and energy dependence parameter $\alpha$=0.005 MeV$^{-1}$, is provided. The magnitude of the incompressibility $K_s$ decreases with the isospin asymmetry $X$ due to lowering of the saturation densities $\rho_s$ with $X$ as well as decrease in the EoS curvature. At high isospin asymmetry $X$, the isospin asymmetric nuclear matter does not have a minimum signifying that it can never be bound by itself due to nuclear interaction. However, the $\beta$ equlibrated nuclear matter which is a highly neutron rich asymmetric nuclear matter exists in the core of the neutron stars since its E/A is lower than that of SNM at high densities and is unbound by the nuclear force but can be bound due to high gravitational field realizable inside neutron stars.      

\noindent 
\begin{table}
\centering
\caption{Incompressibility of isospin asymmetric nuclear matter using the usual value of $n$=2/3 and energy dependence  parameter $\alpha$=0.005 MeV$^{-1}$.}
\begin{tabular}{ccc}
\hline
\hline
$X$&$\rho_s$& $K_s$      \\
\hline
 & fm$^{-3}$ &MeV    \\ 
\hline
 0.0&0.1533&274.7 \\ 
 0.1&0.1525&270.4 \\ 
 0.2&0.1500&257.7 \\ 
 0.3&0.1457&236.6 \\ 
 0.4&0.1392&207.6 \\ 
 0.5&0.1300&171.2 \\  \hline
\hline
\end{tabular} 
\end{table}
\nopagebreak

\vspace{-0.8cm}
      The theoretical estimates of the pressure $P$ and velocity of sound $v_s$ of SNM and isospin asymmetric nuclear matter including PNM are calculated as functions of nucleonic density $\rho$ and energy density $\varepsilon$ using the usual value of 0.005 MeV$^{-1}$ for the parameter $\alpha$ of energy dependence of the zero range potential and also the standard value of the parameter $n$=2/3. Unlike other non-relativistic EoS, present EoS does not suffer from superluminosity at all for SNM and for PNM problem of super luminosity occurs only at very high densities ($\rho>12\rho_0$), higher than those encountered at the centres of neutron stars. In our earlier calculations for the nuclear EoS where the energy dependence of the zero range potential was treated as fixed at a value corresponding to the equilibrium energy per nucleon $\epsilon_0$ \cite{CB06} and assumed to vary negligibly with $\epsilon$ inside nuclear fluid caused superluminosity problems for both SNM and PNM and at much lower densities \cite{BCS06} like the EoS obtained using the $v_{14}+TNI$ interaction \cite{Fr81}. In the present calculations the energy variation of the zero range potential is treated more accurately allowing it to vary freely but only with the kinetic energy part $\epsilon^{kin}$ of the energy per nucleon $\epsilon$ over the entire range of $\epsilon$. 

      In Fig.-1 the energy per nucleon $\epsilon$ of SNM and PNM are plotted as functions of $\rho$. The continuous lines represent the curves for the present calculations using saturation energy per nucleon of -15.26 MeV whereas the dotted lines represent the same using $v_{14}+TNI$ interaction \cite{Fr81} and the dash-dotted lines represent the same for the A18 model using variational chain summation (VCS) \cite{Ak98} for the SNM and PNM. The minimum of the energy per nucleon equaling the saturation energy -15.26 MeV for the present calculations occurs precisely at the saturation density $\rho_0$=0.1533 fm$^{-3}$ since equilibrium density $\rho_0$ of the cold SNM is determined from the saturation condition $\frac{\partial\epsilon}{\partial\rho}$=0 at $\rho$=$\rho_0$ and $\epsilon$=$\epsilon_0$. Fig.-2 presents the plots of the energy per nucleon $\epsilon$ of nuclear matter with different isospin asymmetry X as functions of $\rho/\rho_0$ for the present calculations. The pressure $P$ of SNM and PNM are plotted in Fig.-3 and Fig.-4 respectively as functions of $\rho/\rho_0$. The continuous lines represent the present calculations whereas the dotted lines represent the same using the A18 model using variational chain summation (VCS) of Akmal et al. \cite{Ak98} for the SNM and PNM. The dash-dotted line of Fig.-3 represents plot of $P$ versus $\rho/\rho_0$ for SNM for RMF using NL3 parameter set \cite{La97} and the area enclosed by the coninuous line corresponds
to the region of pressures consistent with the experimental flow data \cite{Da02}. It is interesting to note that the RMF-NL3 incompressibility for SNM is 271.76 MeV \cite{La99} which about the same as 274.7$\pm$7.4 MeV obtained for the present calculation. The areas enclosed by the continuous and the dashed lines in Fig.-4 correspond to the pressure regions for neutron matter consistent with the experimental flow data after inclusion of the pressures from asymmetry terms with weak (soft NM) and strong (stiff NM) density dependences, respectively \cite{Da02}. In Fig.-5 the velocity of sound $v_s$ in SNM and PNM and the energy density $\varepsilon$ of SNM and PNM for the present calculations are plotted as functions of $\rho/\rho_0$. The continuous lines represent the velocity of sound in units of $10^{-2}$c whereas the dotted lines represent energy density in MeV fm$^{-3}$.

\section{The nuclear symmetry energy}

      The nuclear symmetry energy $E_{sym}(\rho)$ represents a penalty levied on the system as it departs from the symmetric limit of equal number of protons and neutrons and can be defined as the energy required per nucleon to change the SNM to pure neutron matter (PNM) \cite{Kl06}

\begin{equation}
 E_{sym}(\rho)=\epsilon(\rho,1) -\epsilon(\rho,0)
\label{seqn18}
\end{equation}
\noindent
and therefore using Eq.(10) for $X=1$ and Eq.(1), the NSE can be given by

\begin{equation}
 E_{sym}(\rho)= (2^{2/3} - 1)\frac{3}{5}E^0_F(\frac{\rho}{\rho_0})^{2/3}+\frac{C}{2} \rho (1 - \beta\rho^n) J_{v01}
\label{seqn19}
\end{equation}
\noindent
where the Fermi energy $E^0_F=\frac{\hbar^2k_{F_0}^2}{2m}$ for the SNM at ground state. 

\vspace{-0.4cm}
\subsection{Nuclear symmetry energy at high baryonic density}
\vspace{-0.4cm}

The first term of the right hand side is the kinetic energy contribution with density dependence of $a_0\rho^{2/3}$ whereas the second term arising due to nuclear interaction has a density dependence of the form of $a_1\rho+a_2\rho^{n+1}+a_3\rho^{5/3}+a_4\rho^{n+5/3}$ since  $J_{v01}$ is a function of $\epsilon^{kin}$ which varies as $\rho^{2/3}$ and $a_0, a_1, a_2, a_3$ and $a_4$ are constants with respect to the nucleonic density $\rho$ or the parameter $n$. If one uses an alternative definition \cite{Wi88} of $E_{sym}(\rho)= \frac{1}{2} \frac{\partial^2\epsilon(\rho,X)}{\partial{X^2}} \mid_{X=0}$ for the nuclear symmetry energy, only the term $(2^{2/3} - 1)$ of the above equation gets replaced by $5/9$ [which is about five percent less than $(2^{2/3} - 1)$] and reduces only the kinetic energy contribution.

      In Fig.-6 plots of E/A for SNM, PNM and NSE as functions of $\rho/\rho_0$ are shown for $n$=2/3. The density dependence of the NSE at subnormal density from isospin diffusion \cite{Ch05} in heavy-ion collisions at intermediate energies has an approximate form of $31.6[\frac{\rho}{\rho_0}]^{1.05}$ MeV. This low energy behaviour of NSE $\approx 31.6[\frac{\rho}{\rho_0}]^{1.05}$ MeV is close to that obtained using Eq.(19) at subnormal densities. At higher densities the present NSE using DDM3Y interaction peaks at $\rho\approx1.95\rho_0$ and becomes negative at $\rho\approx4.7\rho_0$. A negative NSE at high densities implies that the pure neutron matter becomes the most stable state. Consequently, pure neutron matter exists near the core of the neutron stars.  

\vspace{-0.4cm}
\subsection{Nuclear symmetry energy at low baryonic density}
\vspace{-0.4cm}

      The volume symmetry energy coefficient $S_v$ extracted from the masses of finite nuclei provides a constraint on the nuclear symmetry energy at nuclear density $E_{sym}(\rho_0)$. The value of $S_v=30.048 \pm 0.004$ MeV recently extracted \cite{Mu06} from the measured atomic mass excesses of 2228 nuclei is reasonably close to the theoretical estimate of the value of NSE at the saturation density $E_{sym}(\rho_0)$=30.71$\pm$0.26 MeV obtained from the present calculations using DDM3Y interaction. Instead of Eq.(18), if an alternative definition \cite{Wi88} $E_{sym}(\rho)= \frac{1}{2} \frac{\partial^2\epsilon(\rho,X)}{\partial{X^2}} \mid_{X=0}$ of the nuclear symmetry energy is used, then its value is 30.03$\pm$0.26 MeV. In ref. \cite{Da03} it is between 29.10 MeV to 32.67 MeV and that obtained by the liquid droplet model calculation of ref. \cite{St05} is 27.3 MeV whereas in ref. \cite{Di05} it is 28.0 MeV. It should be mentioned that the value of the volume symmetry parameter $S_v$ in some advanced mass description \cite{Po03} is close to the present value which with their $-\kappa_{vol}.b_{vol}=S_v$ equals 29.3 MeV. The value of NSE at nuclear saturation density $\approx$ 30 MeV, therefore, seems well established empirically. Theoretically different parametrizations of the relativistic mean-field (RMF) models, which fit obseravables for isospin symmetric nuclei well, lead to a relatively wide range of predictions 24-40 MeV for $E_{sym}(\rho_0)$. The present result of 30.71$\pm$0.26 MeV of the mean field calculation is close to the results of the calculation using Skyrme interaction SkMP (29.9 MeV) \cite{Be89}, Av18+$\delta v$+UIX$^*$ variational calculation (30.1 MeV) \cite{Ak98} and field theoretical calculation DD-F (31.6 MeV) \cite{Kl06}.   

\vspace{-0.4cm}
\subsection{Compact star constraints}               
\vspace{-0.4cm}

      The knowledge of the density dependence of nuclear symmetry energy is important for understanding not only the stucture of radioactive nuclei but also many important issues in nuclear astrophysics, such as nucleosynthesis during presupernova evolution of massive stars and the cooling of protoneutron stars. A neutron star without neutrino trappings can be considered as a $n,p,e$ matter consisting of neutrons (n), protons (p) and electrons (e). The neutrinos do not accumulate in neutron stars because of its very small interaction probability and correspondingly very high mean free path \cite{Sh03,Je03}. The $\beta$ equilibrium proton fraction $x_\beta$ [$=\rho_p/\rho$] is determined by \cite{La91} 

\begin{equation}
 \hbar c (3 \pi^2\rho x_\beta)^{1/3}= 4E_{sym}(\rho) (1 - 2 x_\beta). 
\label{seqn20}
\end{equation}
\noindent
The equilibrium proton fraction is therefore entirely determined by the NSE. The $\beta$ equilibrium proton fraction calculated using the present NSE is plotted as function of $\rho/\rho_0$ in Fig.-7. The maximum of $x_\beta\approx0.044 $ occurs at $\rho\approx1.35\rho_0$ and goes to zero at $\rho \approx 4.5\rho_0$ for $n$=2/3. The NSE extracted from the isospin diffusion in the intermediate energy heavy-ion collisions, having the approximate form of $31.6[\frac{\rho}{\rho_0}]^{1.05}$ MeV, provides a monotonically increasing $\beta$ equilibrium proton fraction and therefore can not be extended beyond normal nuclear matter densities. Present calculation, using NSE given by Eq.(19), of the $\beta$ equilibrium proton fraction forbids the direct URCA process since the equilibrium proton fraction is always less than 1/9 \cite{La91} which is consistent with the fact that there are no strong indications that fast cooling occurs. Moreover, recently it has been concluded theoretically that an acceptable EoS of asymmetric nuclear matter (such as $\beta$ equilibrated neutron matter) shall not allow the direct URCA process to occur in neutron stars with masses below 1.5 solar masses \cite{Kl06}. Although a recent experimental observation suggests high heat conductivity and enhanced core cooling process indicating the enhanced level of neutrino emission but that can be via the direct URCA process or Cooper-pairing \cite{Ca02}. Also observations of massive compact stars in the mass range of 2.1$\pm$0.2 solar mass to a 1$\sigma$ confidence level (and 2.1$^{+0.4}_{-0.5}$ solar mass to a 2$\sigma$ confidence level) and 2.0$\pm$0.1 solar mass and the lower bound for the mass-radius relation of isolated pulsar RX J1856 imply a rather `stiff' nuclear EoS \cite{Kl06}. The present NSE is `soft' because it increases initially with nucleonic density up to about two times the normal nuclear density and then decreases monotonically at higher densities. It is interesting to observe that although the SNM incompressibility is slightly on the higher side and the present EoS is `stiff', yet the present calculations provide a rather `soft' nuclear symmetry energy and thus satisfy the astrophysical constraints.  

\noindent 
\begin{table}
\caption{Comparison between experimentally measured and theoretically calculated half-lives of spherical proton emitters. The asterisk symbol (*) in the experimental $Q$ values denotes the isomeric state. The experimental $Q$ values, half lives and $l$ values are from ref. \cite{So02}. The results of the present calculations using the isoscalar and isovector components of DDM3Y folded potentials are compared with the experimental values and with the results of UFM estimates \cite{Bal05}. Experimental errors in $Q$ \cite{So02} values and corresponding errors in  calculated half-lives are inside parentheses.}
\begin{tabular}{lllllllll}
\hline
\hline
Parent & $l$ & $Q$ &1$^{st}$ tpt &2$^{nd}$ tpt &3$^{rd}$ tpt &Expt. &This work&UFM       \\ 
$^A Z$& $\hbar$ & MeV &$R_1$[fm]&$R_a$[fm]&$R_b$[fm]&$log_{10}T(s)$&$log_{10}T(s)$& $log_{10}T(s)$ \\ 
\hline
$^{105}Sb$&2&0.491(15)&1.43&6.69&134.30&2.049$^{+0.058}_{-0.067}$&1.90(45)&2.085\\ 
$^{145}Tm$&5&1.753(10)&3.20&6.63&56.27&-5.409$^{+0.109}_{-0.146}$&-5.28(7)&-5.170\\ 
$^{147}Tm$&5&1.071(3)&3.18&6.63&88.65&0.591$^{+0.125}_{-0.175}$&0.83(4)&1.095\\ 
$^{147}Tm^*$&2&1.139(5)&1.44&7.28&78.97&-3.444$^{+0.046}_{-0.051}$&-3.46(6)&-3.199\\ 
$^{150}Lu$&5&1.283(4)&3.21&6.67&78.23&-1.180$^{+0.055}_{-0.064}$&-0.74(4)&-0.859\\ 
$^{150}Lu^*$&2&1.317(15)&1.45&7.33&71.79&-4.523$^{+0.620}_{-0.301}$&-4.46(15)&-4.556\\ 
$^{151}Lu$&5&1.255(3)&3.21&6.69&78.41&-0.896$^{+0.011}_{-0.012}$&-0.82(4)&-0.573\\ 
$^{151}Lu^*$&2&1.332(10)&1.46&7.35&69.63&-4.796$^{+0.026}_{-0.027}$&-4.96(10)&-4.715\\ 
$^{155}Ta$&5&1.791(10)&3.21&6.78&57.83&-4.921$^{+0.125}_{-0.125}$&-4.80(7)&-4.637\\ 
$^{156}Ta$&2&1.028(5)&1.47&7.37&94.18&-0.620$^{+0.082}_{-0.101}$&-0.47(8)&-0.461\\ 
$^{156}Ta^*$&5&1.130(8)&3.21&6.76&90.30&0.949$^{+0.100}_{-0.129}$&1.50(10)&1.446\\ 
$^{157}Ta$&0&0.947(7)&0.00&7.55&98.95&-0.523$^{+0.135}_{-0.198}$&-0.51(12)&-0.126\\ 
$^{160}Re$&2&1.284(6)&1.45&7.43&77.67&-3.046$^{+0.075}_{-0.056}$&-3.08(7)&-3.109\\ 
$^{161}Re$&0&1.214(6)&0.00&7.62&79.33&-3.432$^{+0.045}_{-0.049}$&-3.53(7)&-3.231\\ 
$^{161}Re^*$&5&1.338(7)&3.22&6.84&77.47&-0.488$^{+0.056}_{-0.065}$&-0.75(8)&-0.458\\ 
$^{164}Ir$&5&1.844(9)&3.20&6.91&59.97&-3.959$^{+0.190}_{-0.139}$&-4.08(6)&-4.193\\ 
$^{165}Ir^*$&5&1.733(7)&3.21&6.93&62.35&-3.469$^{+0.082}_{-0.100}$&-3.67(5)&-3.428\\ 
$^{166}Ir$&2&1.168(8)&1.47&7.49&87.51&-0.824$^{+0.166}_{-0.273}$&-1.19(10)&-1.160\\ 
$^{166}Ir^*$&5&1.340(8)&3.22&6.91&80.67&-0.076$^{+0.125}_{-0.176}$&0.06(9)&0.021\\ 
$^{167}Ir$&0&1.086(6)&0.00&7.68&91.08&-0.959$^{+0.024}_{-0.025}$&-1.35(8)&-0.943\\ 
$^{167}Ir^*$&5&1.261(7)&3.22&6.92&83.82&0.875$^{+0.098}_{-0.127}$&0.54(8)&0.890\\ 
$^{171}Au$&0&1.469(17)&0.00&7.74&69.09&-4.770$^{+0.185}_{-0.151}$&-5.10(16)&-4.794\\ 
$^{171}Au^*$&5&1.718(6)&3.21&7.01&64.25&-2.654$^{+0.054}_{-0.060}$&-3.19(5)&-2.917\\ 
$^{177}Tl$&0&1.180(20)&0.00&7.76&88.25&-1.174$^{+0.191}_{-0.349}$&-1.44(26)&-0.993\\ 
$^{177}Tl^*$&5&1.986(10)&3.22&7.10&57.43&-3.347$^{+0.095}_{-0.122}$&-4.64(6)&-4.379\\ 
$^{185}Bi$&0&1.624(16)&0.00&7.91&65.71&-4.229$^{+0.068}_{-0.081}$&-5.53(14)&-5.184\\ \hline
\hline
\end{tabular} 
\end{table}
\noindent

\vspace{-0.8cm}

\section{Folding model analyses using effective interaction whose density dependence determined from nuclear matter calculation}

      Microscopic proton-nucleus interaction potentials are obtained by single folding the density of the nucleus with M3Y effective interaction supplemented by a zero-range pseudo-potential for exchange along with the density dependence. Parameters of the density dependence, $C$=2.2497 and $\beta$=1.5934 fm$^2$, obtained here from the nuclear matter calculations assuming kinetic energy dependence of zero range potential, are used. 

      The half lives of the decays of spherical nuclei away from proton drip line by proton emissions are estimated theoretically. The half life of a parent nucleus decaying via proton emission is calculated using the WKB barrier penetration probability. The WKB method is found quite satisfactory and even better than the S-matrix method for calculating half widths of the $\alpha$ decay of superheavy elements \cite{Ma06}. For the present calculations, the zero point vibration energies used here are given by eqn.(5) of ref. \cite{Po86} extended to protons and the experimental $Q$ values \cite{So02} are used. Spherical charge distributions are used for Coulomb interaction potentials. The same set of data of ref. \cite{Bal05} has been used for the present calculations using $C$=2.2497 and $\beta$=1.5934 fm$^2$ and presented in Table-2. The agreement of the present calculations with a wide range of experimental data for the proton radioactivity lifetimes are reasonable. 

      Since the density dependence of the effective projectile-nucleon interaction was found to be fairly independent of the projectile \cite{Sr83}, as long as the projectile-nucleus interaction was amenable to a single-folding prescription, the density dependent effects on the nucleon-nucleon interaction were factorized into a target term times a projectile term and used successfully in case of $\alpha$ radioctivity of nuclei \cite{Ba03} including superheavies \cite{CSB06,prc07,scb07} and the cluster radioactivity \cite{Ba03,Bas02}. The calculations were performed for elastic and inelastic scattering of protons from nuclei $^{18}Ne$, $^{18}O$, $^{20}O$, $^{22}O$ using $C$=2.07 and $\beta$=1.624 fm$^2$ \cite{Gu05,Gu06}. It is needless to say that the present value of $\beta$=1.5934$\pm$0.0085 fm$^2$, obtained by treating the energy variation of the zero range potential properly by allowing it to vary freely with the kinetic energy, which changes by about one percent neither changes the shape of the potential significantly nor the quality of fit or the values extracted for the nuclear deformations. However, since the value of $C$, which acts as the overall normalisation constant for the nuclear potentials, changes by about six percent, causes changes but only to the renormalizations required for the potentials. Therefore, it provides reasonable description for elastic and inelastic scattering of protons and the deformation parameters extracted from these analyses are in good agreement with the quadrupole deformations obtained from the available experimental $B(E2)$ values \cite{RA87}.  

\vspace{-0.25cm}
\section{Summary and conclusion}
\vspace{-0.25cm}

      A mean field calculation is carried out to obtain the equation of state of nuclear matter from a density dependent M3Y interaction (DDM3Y). The microscopic nuclear potentials are obtained by folding the DDM3Y effective interaction with the densities of interacting nuclei. The energy per nucleon is minimized to obtain ground state of the symmetric nuclear matter (SNM). The constants of density dependence of the effective interaction are obtained by reproducing the saturation energy per nucleon and the saturation density of SNM. The EoS of asymmetric nuclear matter is calculated by adding to the isoscalar part, the isovector component of M3Y interaction. The SNM and pure neutron matter EoS are used to calculate the nuclear symmetry energy which is found to be consistent with that extracted from the isospin diffusion in heavy-ion collisions at intermediate energies. The microscopic proton-nucleus interaction potential is obtained by folding the density of the nucleus with DDM3Y effective interaction whose density dependence is determined completely from the nuclear matter calculations. 

      In this work the energy variation of the exchange potential is treated properly in the negative energy domain of nuclear matter. The EoS of SNM, thus obtained, is free from the superluminosity problem encountered in some previous prescriptions. Moreover, the result of the present calculation for the compression modulus for the infinite symmetric nucler matter is in better agreement with that extracted from experiments. The calculated $\beta$ equilibrium proton fraction forbids direct URCA process which is consistent with the fact that there are no strong indications that fast cooling occurs. The results of the present calculations using single folded microscopic potentials for the proton-radioactivity lifetimes are in good agreement over a wide range of experimental data. We find that it also provides reasonable description for the elastic and inelastic scattering of protons and the deformation parameters extracted from the analyses are in good agreement with the available results. The results of the present calculations using microscopic potentials for half life calculations of $\alpha$ decays are found to be in excellent agreement with experimental data. These calculations also provide reliable estimates for the observed $\alpha$ decay lifetimes of the newly synthesized superheavy elements. It is, therefore, pertinent to conclude that a unified description of the symmetric and asymmetric nuclear matter, elastic and inelastic scattering, and cluster, $\alpha$ and proton radioactivities is achieved. With the energies and interaction rates foreseen at FAIR, the compressed baryonic matter (CBM) will create highest baryon densities in nucleus-nucleus collisions to explore the properties of superdense baryonic matter and the in-medium modifications of hadrons.

\pagebreak

\begin{figure}[h]
\eject\centerline{\epsfig{file=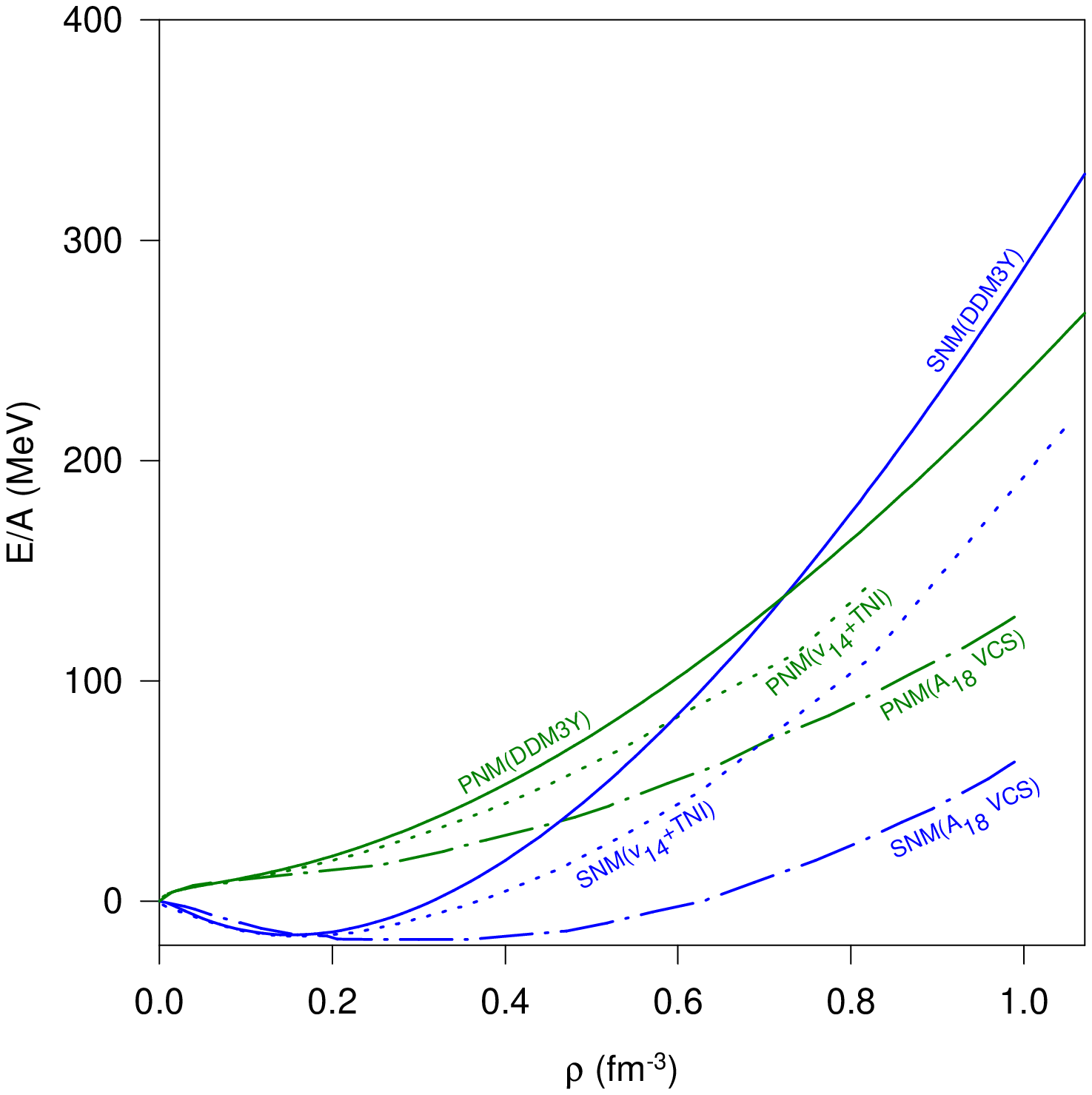,height=14cm,width=14cm}}
\caption
{The energy per nucleon $\epsilon$ = E/A of SNM (spin and isospin symmetric nuclear matter) and PNM (pure neutron matter) as functions of $\rho$. The continuous lines represent curves for the present calculations using DDM3Y interaction, the dotted lines represent the same using $v_{14}+TNI$ interaction \cite{Fr81} and the dash-dotted lines represent the same for A18 model using variational chain summation (VCS) \cite{Ak98}.}
\label{fig1}
\end{figure}

\begin{figure}[h]
\eject\centerline{\epsfig{file=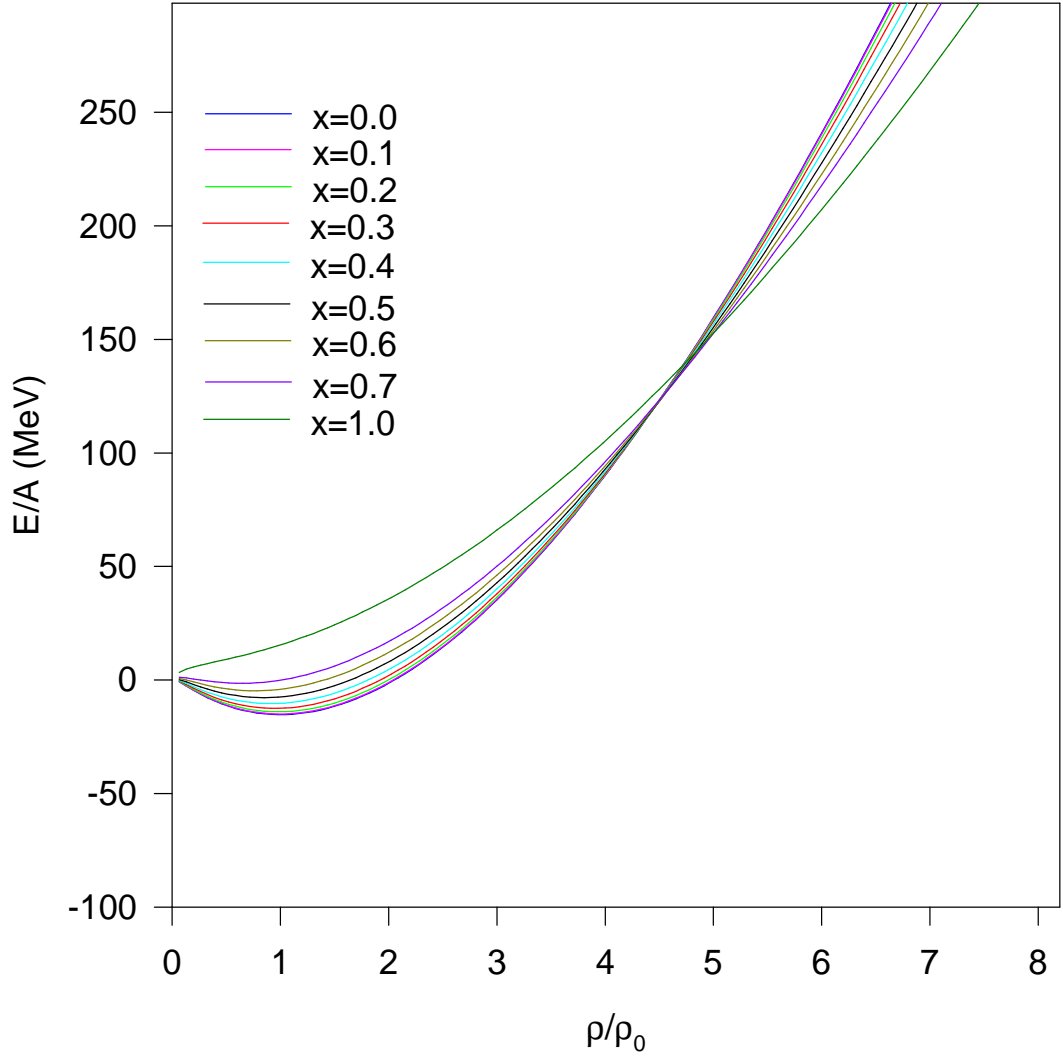,height=14cm,width=14cm}}
\caption
{The energy per nucleon $\epsilon$=E/A of nuclear matter with different isospin asymmetry X as functions of $\rho/\rho_0$ for the present calculations using DDM3Y interaction.}
\label{fig2}
\end{figure}

\begin{figure}[h]
\eject\centerline{\epsfig{file=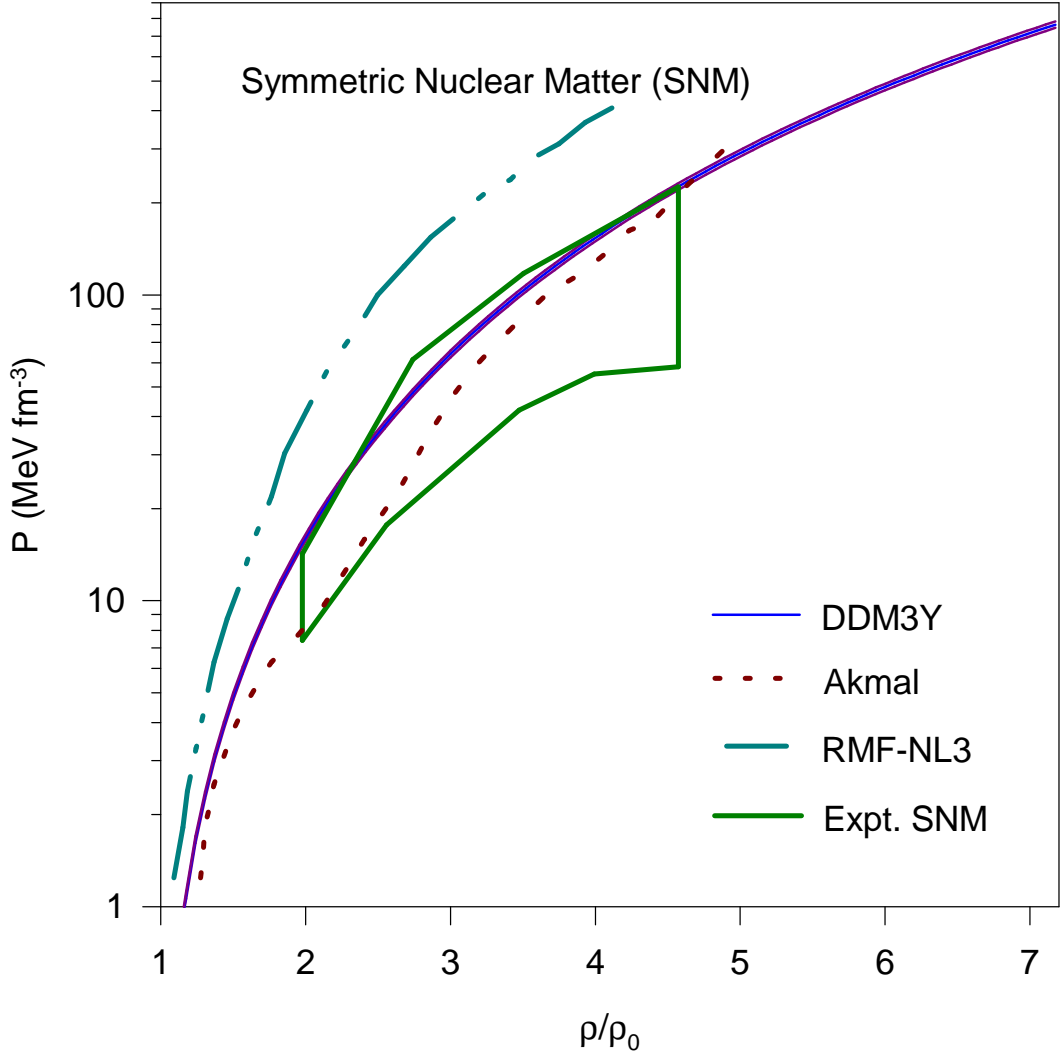,height=14cm,width=14cm}}
\caption
{The pressure $P$ of SNM (spin and isospin symmetric nuclear matter) as a function of $\rho/\rho_0$. The continuous lines represent the present calculations using $\epsilon_0$ = -15.26$\pm$0.52 MeV. The dotted line represents the same using the A18 model using variational chain summation (VCS) of Akmal et al. \cite{Ak98}, the dash-dotted line represents the RMF calculations using NL3 parameter set \cite{La97} whereas the area enclosed by the coninuous line corresponds
to the region of pressures consistent with the experimental flow data \cite{Da02} for SNM.}
\label{fig3}
\end{figure}

\begin{figure}[h]
\eject\centerline{\epsfig{file=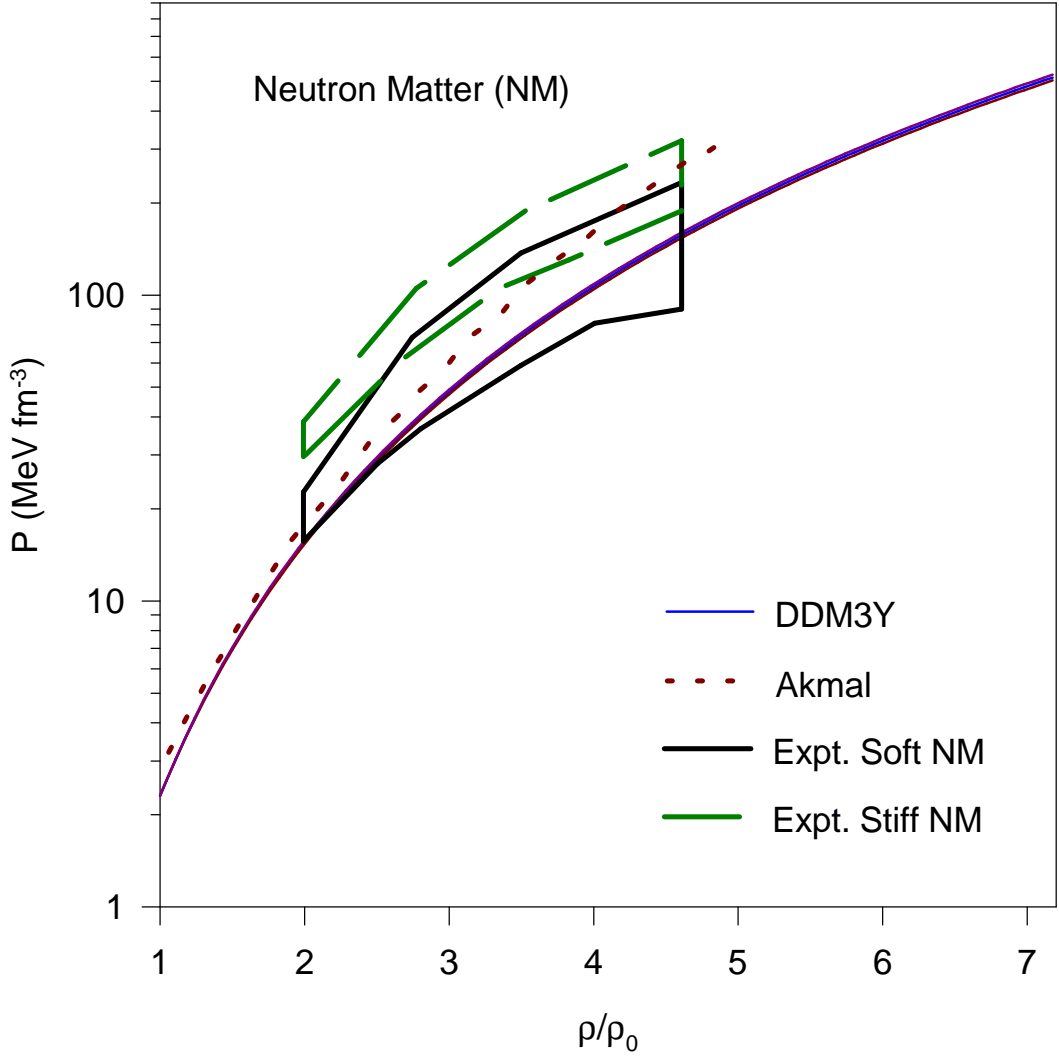,height=14cm,width=14cm}}
\caption
{The pressure $P$ of PNM (pure neutron matter) as a function of $\rho/\rho_0$. The continuous lines represent the present calculations using $\epsilon_0$ = -15.26$\pm$0.52 MeV. The dotted line represents the same using the A18 model using variational chain summation (VCS) of Akmal et al. \cite{Ak98} whereas the areas enclosed by the continuous and the dashed lines correspond to the pressure regions for neutron matter consistent with the experimental flow data after inclusion of the pressures from asymmetry terms with weak (soft NM) and strong (stiff NM) density dependences, respectively \cite{Da02}.}
\label{fig4}
\end{figure}

\begin{figure}[h]
\eject\centerline{\epsfig{file=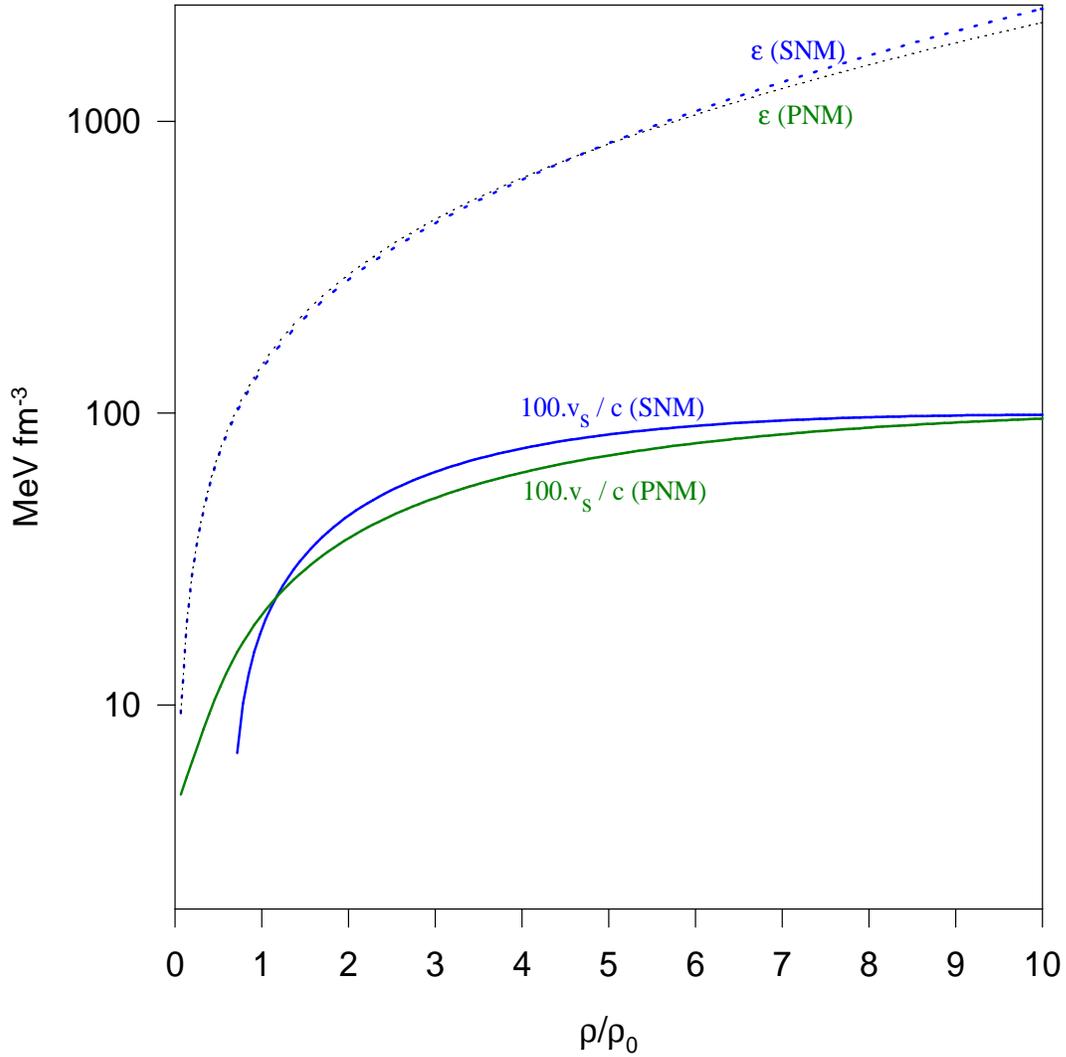,height=14cm,width=14cm}}
\caption
{The velocity of sound $v_s$ in SNM (spin and isospin symmetric nuclear matter) and PNM (pure neutron matter) and the energy density $\varepsilon$ of SNM and PNM as functions of $\rho/\rho_0$ for the present calculations using DDM3Y interaction. The continuous lines represent the velocity of sound in units of $10^{-2}c$ whereas the dotted lines represent energy density in MeV fm$^{-3}$.}
\label{fig5}
\end{figure}

\begin{figure}[h]
\eject\centerline{\epsfig{file=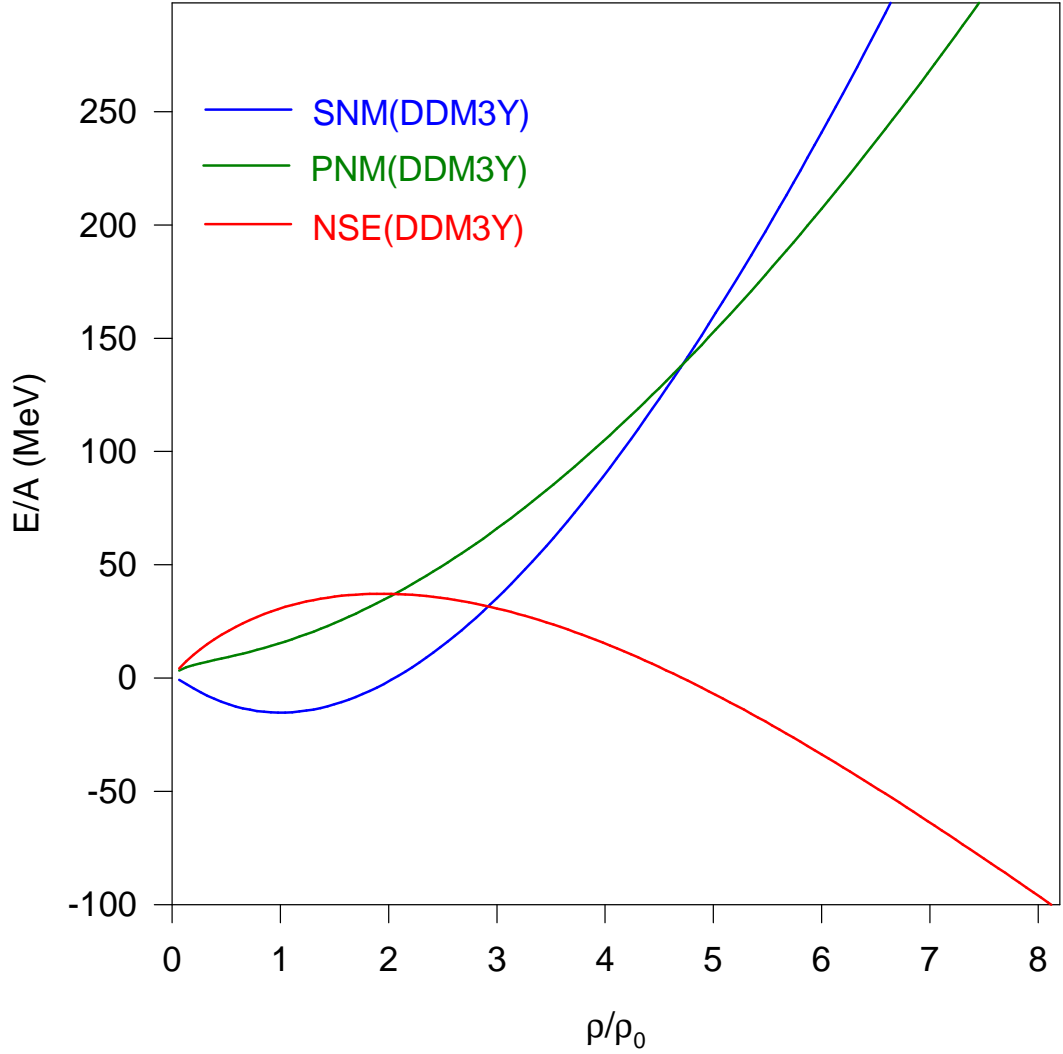,height=14cm,width=14cm}}
\caption
{The energy per nucleon $\epsilon$=E/A of SNM (spin and isospin symmetric nuclear matter), PNM (pure neutron matter) and NSE (nuclear symmetry energy $E_{sym}$) are plotted as functions of $\rho/\rho_0$ for the present calculations using DDM3Y interaction.}
\label{fig6}
\end{figure}

\begin{figure}[h]
\eject\centerline{\epsfig{file=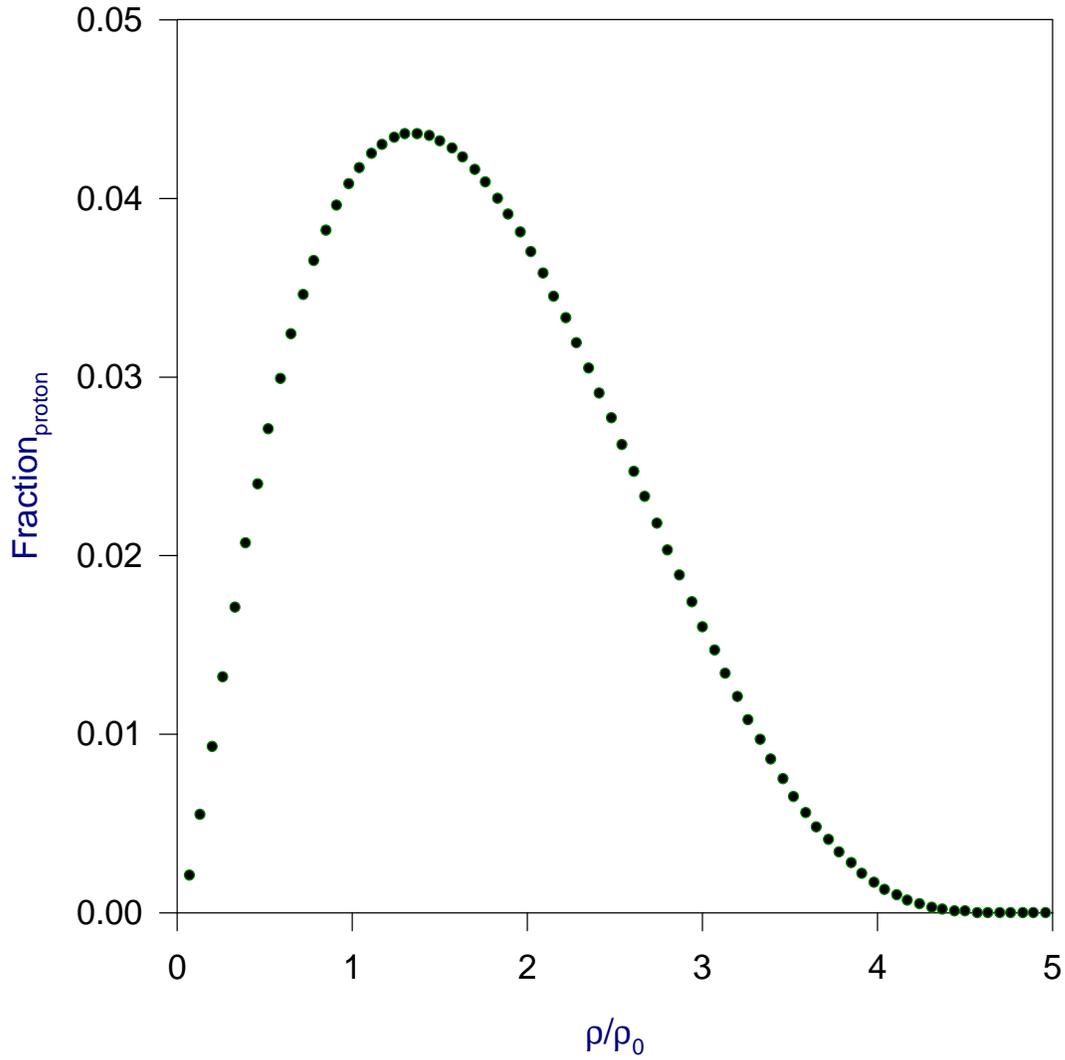,height=14cm,width=14cm}}
\caption
{The $\beta$ equilibrium proton fraction calculated with NSE (nuclear symmetry energy) obtained using DDM3Y interaction is plotted as a function of $\rho/\rho_0$.}
\label{fig7}
\end{figure}
                                   
\end{document}